\documentclass[preprint2]{aastex631}

\usepackage{amsmath,amssymb}

\graphicspath{{./}{figures/}}

\begin{document}

\title{Statistical analysis of intermittency and its association with proton heating in the near Sun environment.}


\author[0000-0002-1128-9685]{Nikos Sioulas}
\affiliation{Earth, Planetary, and Space Sciences, \;
University of California, Los Angeles \\
Los Angeles, CA 90095, USA }

\author[0000-0002-2381-3106]{Marco Velli}
\affiliation{Earth, Planetary, and Space Sciences, \;
University of California, Los Angeles \\
Los Angeles, CA 90095, USA }

\author[0000-0002-7174-6948]{Rohit Chhiber}
\affiliation{Department of Physics and Astronomy, Bartol Research Institute, University of Delaware \\  Newark, DE 19716, USA}
\affiliation{ NASA Goddard Space Flight Center,  \\ Greenbelt,   MD 20771, USA}


\author[0000-0002-8700-4172]{Loukas Vlahos}
\affiliation{Deprtment of Physics, \; Aristotle University of Thessaloniki\\
GR-52124 Thessaloniki, Greece }

\author[0000-0001-7224-6024]{William H. Matthaeus}
\affiliation{Department of Physics and Astronomy, Bartol Research Institute, University of Delaware \\  Newark, DE 19716, USA}

\author[0000-0002-6962-0959]{Riddhi Bandyopadhyay}
\affiliation{ Department of Astrophysical Sciences, Princeton University, \\
 Princeton, NJ 08544, USA}

\author[0000-0002-7341-2992]{Manuel E. Cuesta}
\affiliation{Department of Physics and Astronomy, Bartol Research Institute, University of Delaware \\  Newark, DE 19716, USA}

\author[0000-0002-2582-7085]{Chen Shi}
\affiliation{Earth, Planetary, and Space Sciences, \;
University of California, Los Angeles \\
Los Angeles, CA 90095, USA }

\author[0000-0002-4625-3332]{ Trevor A. Bowen}
\affiliation{Space Sciences Laboratory, University of California, \\ Berkeley, CA 94720-7450, USA}

\author[0000-0001-8358-0482]{Ramiz A. Qudsi}
\affiliation{Department of Physics and Astronomy, Bartol Research Institute, University of Delaware, \\  Newark, DE 19716, USA}

\author[0000-0002-7728-0085]{Michael L. Stevens}
\affiliation{Harvard-Smithsonian Center for Astrophysics, \\ Cambridge, MA 02138, USA,}

\author[0000-0002-1989-3596]{Stuart D. Bale }
\affiliation{Space Sciences Laboratory, University of California, \\ Berkeley, CA 94720-7450, USA}
\affiliation{Physics Department, University of California, \\ Berkeley, CA 94720-7300, USA}

\date{\today}
\begin{abstract}

We use data from the first six encounters of Parker Solar Probe and employ the Partial Variance of Increments ($PVI$) method to study the statistical properties of coherent structures in the inner heliosphere with the aim of exploring physical connections between magnetic field intermittency and observable consequences such as plasma heating and turbulence dissipation. Our results support proton heating localized in the vicinity of, and strongly correlated with, magnetic structures characterized by $PVI \geq 1$. We show that on average, such events constitute $\approx 19\%$ of the dataset, though variations may occur depending on the plasma parameters. We show that the waiting time distribution ($WT$) of identified events is consistent across all six encounters following a power-law scaling at lower $WTs$. This result indicates that coherent structures are not evenly distributed in the solar wind but rather tend to be tightly correlated and form clusters. We observe that the strongest magnetic discontinuities, $PVI \geq 6$, usually associated with reconnection exhausts, are sites where magnetic energy is locally dissipated in proton heating and are associated with the most abrupt changes in proton temperature. However, due to the scarcity of such events, their relative contribution to energy dissipation is minor. Taking clustering effects into consideration, we show that smaller scale, more frequent structures with  PVI between, $1\lesssim PVI \lesssim 6$, play the major role in magnetic energy dissipation.  The number density of such events is strongly associated with the global solar wind temperature, with denser intervals being associated with higher $T_{p}$.

\end{abstract}
\clearpage

\section{Introduction}\label{sec:intro}

The solar wind is a strongly magnetized, weakly collisional stream of charged particles expanding at supersonic speeds from the outermost layer of the sun, the corona \citep{1958ApJ...128..664P}.
For coronal temperatures of order $\sim 100 \ eV $ and adiabatic cooling, the temperature at 1 AU would be expected to decrease to below  $\sim 1 \ eV $. Observations of the solar wind, however, show values in the order of $\sim 10 \ eV $ at 1 AU \citep{1994JGR....99.6561G, 1995GeoRL..22..325R} and ion temperature decays as a function of radial distance as $T \sim r^{- \gamma}$, with the exponent, attaining values in the range, $0.5\lesssim \gamma \lesssim 1$ \citep{https://doi.org/10.1029/94GL03273, 2018SoPh..293..155S}, a much slower rate than spherically symmetric adiabatic expansion models would predict (i.e. $T \sim  r^{-4/3})$. 
A complete description of the dynamics of the solar wind must therefore include non-adiabatic heating processes that contribute to its internal energy. Most theoretical investigations suggest collisionless dissipation mechanisms that can be broadly classified into two categories: wave-particle interactions and current sheets/reconnection
\citep{1989PhRvL..63.1807V, 1999ApJ...523L..93M,  2005ApJS..156..265C,2007ApJS..171..520C, 2014ApJ...796..111L, Isliker19, Sioulas20b, Sioulas20, 2021JPlPh..87b9018P, Gonz_lez_2021}. 
Despite all these efforts, the non-adiabatic expansion of the solar wind remains one of the major outstanding problems in the field of space physics. \par
The turbulent cascade in space plasma systems is a multi-scale process during which the energy entering the system at large scales is removed from the turbulent cascade at the scale of the ion gyroradius (kinetic scales) due to non-linear interactions among the fluctuations \citep{Biskamp03,1968ApJ...153..371C}. This is not the complete story though as, both simulations  \citep{1980NYASA.357..203M, PhysRevLett.47.1060} and observations \citep{https://doi.org/10.1029/JA095iA06p08211,Chhiber_2020} indicate that turbulence also produces intermittency. In the solar wind, fluctuations that extend over several decades in the frequency domain, from the inverse of the solar rotation period down to the electron gyrofrequency, abound.
In this sense, the solar wind is regarded as the exquisite turbulent plasma laboratory, as it offers the chance to study coherent structures, such as current sheets and vortices, being dynamically generated as the end result of the turbulent cascade \citep{Veltri_1999, Greco10}. 

Coherent structures and spatial intermittency arise from the non-self-similarity of the nonlinear cascade, i.e. the probability distribution functions of fluctuations of a field $\phi$, $\Delta \phi = \phi(\ell + \Delta \ell) - \phi(\ell),$ at a given length , $\ell$, display larger and larger tails with respect to a Gaussian distribution as the lengths become smaller and smaller \citep{1990PhyD...46..177C}. This behavior, often described as multifractal, is associated with the occurrence at small scales of coherent structures, superposed on a background of random fluctuations. 
Indeed such structures that persist in time longer than the surrounding stochastic fluctuations are characterized by non-Gaussian statistics and constitute only a minor fraction of the entire dataset \citep{PhysRevLett.108.261102, https://doi.org/10.1029/2018EA000535}. They nevertheless strongly influence the dissipation, heating, transport, and acceleration of charged particles \citep{Matthaeus11, 2013PhPl...20a2303K, Tessein_2013, Bandyopadhyay_2020}. In recent years, several numerical simulations \citep{Biskamp00, 2009PhPl...16c2310P, Lehe_2009, 2012PhRvL.108d5001S, 2021ApJ...919..103J, 2022A&A...657A...8S} have suggested localized, intermittent heating in structures such as reconnection sites and current sheets.

Additionally, observational studies have indicated a statistical link between coherent magnetic field structures and elevated ion temperatures, both in the near-Earth solar wind  \citep{2000ApJ...537.1054L, PhysRevLett.108.261102, 2021ApJ...921...65Y}, as well as in the near-Sun \citep{2020ApJS..246...46Q} environment. As Parker Solar Probe (PSP) \citep{2016SSRv..204....7F} moves closer to the solar wind sources, it offers an unprecedented opportunity to study intermittency and the physics of dissipation in the near sun solar wind environment. 
In this article, we make use of the PVI method and take advantage of the high-resolution magnetic field and particle PSP data from the first six encounters to perform a statistical study of intermittency.  Furthermore, the relationship between solar wind turbulence and the concept of intermittent heating in the nascent solar wind environment is investigated. 

The outline of this work is as follows. Sections \ref{sec:Background}, \ref{sec:Background.2} define the PVI and LET methods respectively, and provide some background; Section \ref{sec:data} presents the selection of PSP data and their processing. The results of our analysis are presented in \ref{sec:results}; in subsection \ref{subsec:Stat_properties} the statistics of discontinuities in the magnetic field obtained by the PVI method are analyzed, while in subsection \ref{subsec:heating}, the effects of the identified discontinuities in the proton temperature of the solar wind are investigated. A summary of the results along with the conclusions is finally given in Section  \ref{sec:Summary}.

\section{Partial Variance of Increments (PVI)}\label{sec:Background}

When confronted with a dataset that samples a turbulent space plasma system, an important task is to find the subset of the data that corresponds to the underlying coherent structures. In recent years, a plethora of methods have been suggested for the identification of intermittent structures and discontinuities in the magnetic field. Some of those include the Phase Coherence Index method \citep{2003SSRv..107..463H}
and the wavelet-based Local Intermittency Measure (LIM) \citep{1999ESASP.448.1147B}. In this paper, we employ a simple and well-studied method that has been effectively used in the past for the study of intermittent turbulence and the identification of coherent structures, both in simulations \citep{Servidio11}, and observations \citep{2018ApJ...862...32C, 2021ApJ...908..237H}, namely, the Partial Variance of Increments (PVI). The advantage of the PVI method is that it provides an easy-to-implement tool that measures the sharpness of a signal relative to the neighborhood of that point. For lag $\tau$, the normalized  $PVI$ at time t is defined as:

\begin{equation}
    PVI( t , \tau) \ = \  \frac{| \Delta\textbf{B}( t, \tau)|}{\sqrt{\langle | \Delta\textbf{B}( t,\tau)|^{2}\rangle}},
\end{equation}

where, $| \Delta\textbf{B}( t, \tau)| \ = \ |\textbf{B}(t \ + \ \tau) \ -  \ \textbf{B}(t)| $ is the magnitude of the magnetic field vector increments, and $\langle...\rangle$ denotes average over a window that is a multiple of the estimated correlation time.  PVI is a threshold method, so to proceed with the analysis, one can impose a threshold, ${\theta}$ on the PVI and select portions in which $PVI > {\theta}$. One could easily see that ${\langle}PVI^{4}{\rangle}$ is related to the kurtosis of the magnetic field, and thus a possible way to calibrate the PVI method is to exclude all the values above ${\theta}$ and compute the kurtosis of the remaining data. The process can be repeated several times, each time lowering the value of ${\theta}$ until the kurtosis of the remaining signal is equal to the value expected for Gaussian random signals. \cite{2018SSRv..214....1G} have shown that increments with $PVI > 3$  lay on the ``heavy tails'' observed in the distribution of increments and can thus be associated with Non-Gaussian structures. By increasing the threshold value $\theta$, one can thus identify the most intense magnetic field discontinuities like current sheets and reconnection sites. Finally, note that the method is insensitive to the mechanism that generates the coherent structures. This means that the PVI can be implemented for the identification of any form of sharp gradients in the vector magnetic field. A more comprehensive review of PVI, as well as a comparison with aforementioned methods, appropriate for identifying discontinuities, can be found in \cite{2018SSRv..214....1G}.

\section{Local Energy Transfer (LET)}
\label{sec:Background.2}

As already noted in Sec. \ref{sec:intro}, coherent structures constitute a minor fraction of the dataset. At the same time coherent structures correspond to sites where significant heating and dissipation takes place, contributing a disproportionate amount to the internal energy of the solar wind. Once the subset of the data-set corresponding to coherent structures has been identified, a method that can allow us to quantify their contribution to energy dissipation and heating of the solar wind is necessary. The Kolmogorov-Yaglom law, extended to isotropic MHD  \citep{politano1998dynamical, PhysRevE.57.R21} (PP98) has been used to evaluate the average energy transfer rate per unit mass, at scale $\Delta t$. Taking into account the conservation law of the respective inviscid invariants, the scaling law can be directly derived from the dynamical MHD equations for the third-order moment. Assuming homogeneity, scale separation, isotropy, and time-stationarity  PP98 yields the linear scaling of the mixed third-order moment of the field fluctuations

\begin{figure*}
\centering
\setlength\fboxsep{0pt}
\setlength\fboxrule{0.0pt}

\fbox{
         \includegraphics[width=0.45\textwidth]{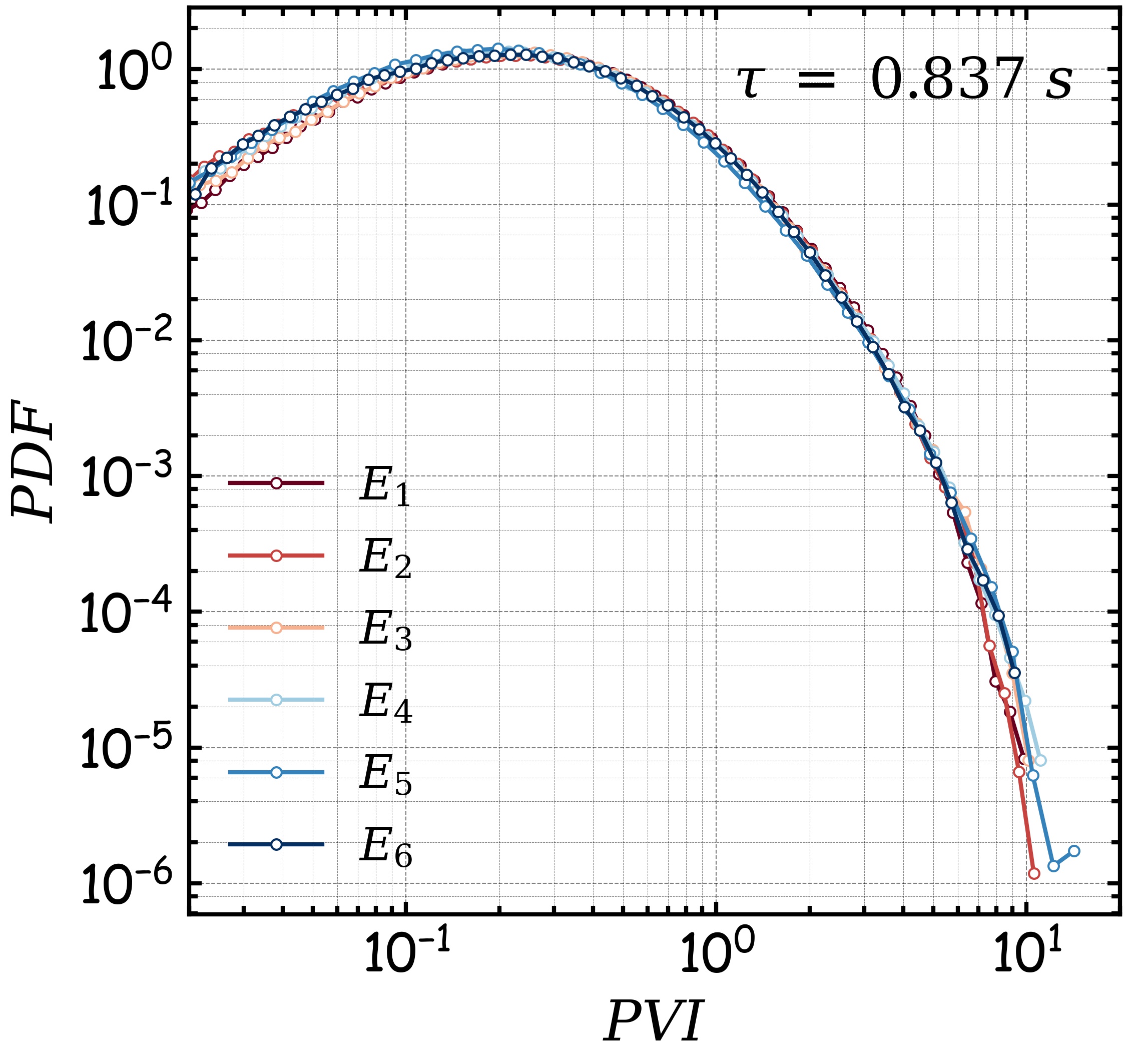}}
\fbox{
         \includegraphics[width=0.45\textwidth]{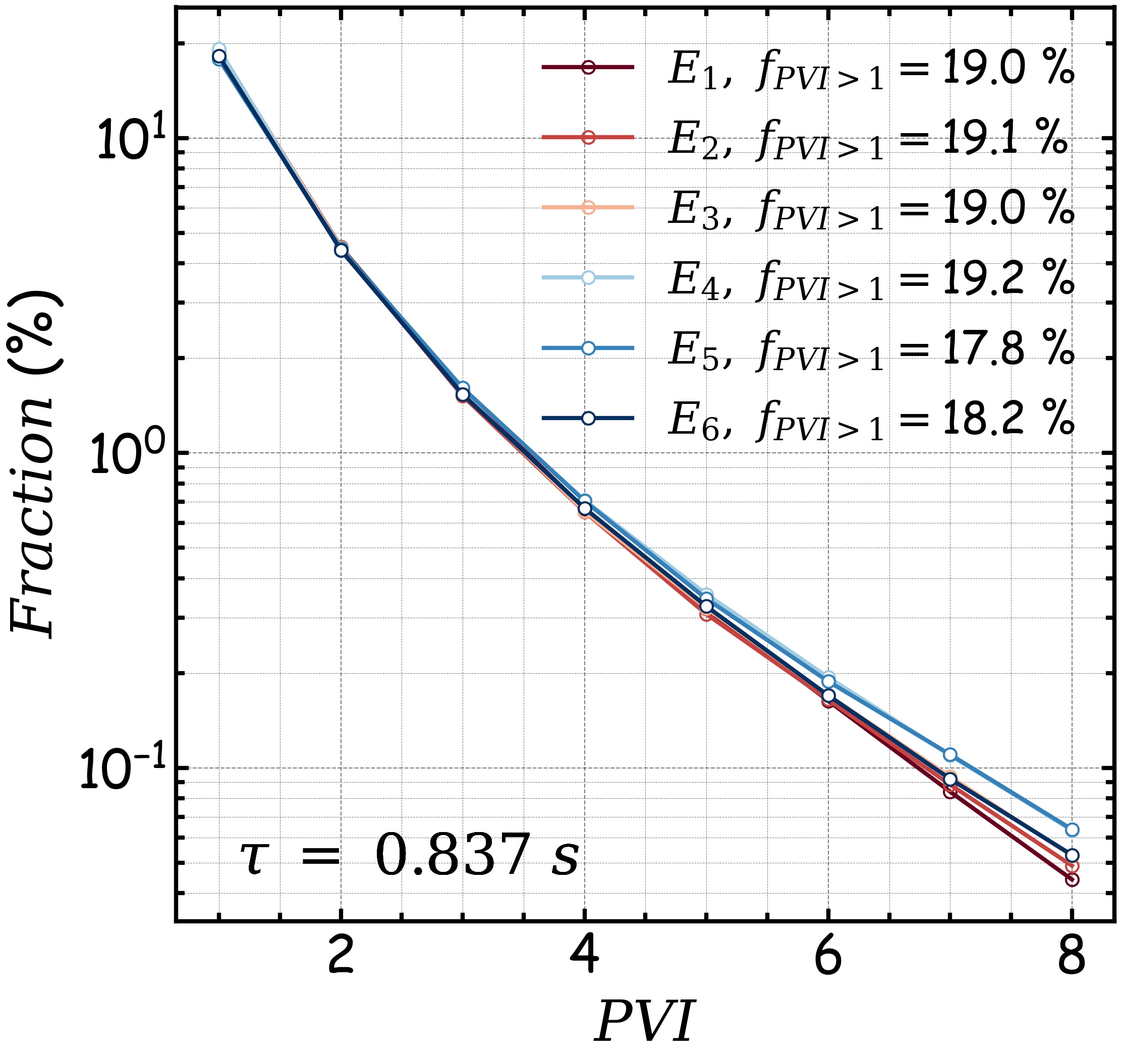}}
\caption{(a) PDFs of $PVI$ for lag,  $\tau \ = 0.837s$  (b) Fraction of PVI events exceeding a given threshold $\theta$ for $E_{1} - E_{6}$.}
\label{fig:PVI1}
\end{figure*}


\begin{multline}\label{eq:PP98}
    Y(\Delta t) = \langle  
    \Delta v_{r}(|\Delta \textbf{v}|^{2} + |\Delta \textbf{b}|^{2})
    - \Delta b_{r}(\Delta \textbf{v} \cdot \Delta \textbf{b})\rangle  \\ = - \frac{4}{3} \langle \epsilon \rangle  V_{SW} \Delta t,
\end{multline}

where, $\langle ... \rangle$ indicates spatial averaging, and $\langle \epsilon \rangle$ is the mean energy transfer rate. The mixed third-order moment  is computed using the scale-dependent increments of the plasma velocity field \textbf{v}, and the magnetic field given in velocity units, $ \textbf{b} = \frac{\textbf{B}}{\sqrt{\mu_{0} m_{p} n_{p}}} $, where $ \mu _{0}$ is the magnetic permeability of vacuum, $ m_{p}$
is the proton mass, and $n_p$ is the number density of
protons. The subscript r indicates the longitudinal component, i.e., measured in the sampling direction. Note, that the validity of the Taylor's hypothesis \citep{1938RSPSA.164..476T}, has been assumed for converting  spatial to temporal scales through
 \begin{equation}
     \ell  = V_{SW}  \Delta t,
 \end{equation}

The linear relation in Eq. \ref{eq:PP98} has been confirmed under various plasma conditions revealing the turbulent dynamics and providing an estimate of the energy transfer rate of space plasmas throughout the heliosphere \citep{ article, 10.3389/fphy.2019.00108,  2018ApJ...866..106B, 10150/657358,2021MNRAS.500L...6B,2021ApJ...922L..11H}.
 In recent years, it was suggested \citep{article, 10.3389/fphy.2019.00108} that by omitting the averaging process in Eq. \ref{eq:PP98}, a proxy  of the local energy transfer rates (LET) at a given scale $\Delta t$  can be obtained as

 \begin{equation}
 \small{\epsilon^{\pm}(t, \Delta t) = -\frac{3}{4} \frac{\Delta v_{r}(|\Delta \textbf{v}|^{2} + |\Delta \textbf{b}|^{2})
    - \Delta b_{r}(\Delta \textbf{v} \cdot \Delta \textbf{b})}{V_{SW} \Delta t}}.
\end{equation}

The LET is composed of two additive terms. The first term, $\epsilon^{e}=-3/(4 V_{SW} \Delta t ) [\Delta v_{r}(|\Delta \textbf{v}|^{2} + |\Delta \textbf{b}|^{2})$], is associated with the magnetic and 
kinetic energy advected by the velocity fluctuations. \ The  \ second \ \  term, \ \  \ \    $\epsilon^{c} =$-$   3/(4 V_{SW} \Delta t)[\Delta b_{r}(\Delta \textbf{v} \cdot \Delta \textbf{b})]$, is associated with the cross-helicity coupled to the longitudinal magnetic fluctuations. Note that several important scaling contributions are neglected in this procedure of estimating LET. Such contributions in Eq. \ref{eq:PP98}, are suppressed by averaging over a large sample. Consequently, this approach provides only a
local proxy, indicating the contribution to the local  energy transfer rate \citep{PhysRevE.99.053202}
 \vspace{0.6cm}

\begin{figure*}
\centering
\setlength\fboxsep{0pt}
\setlength\fboxrule{0.0pt}
\fbox{\includegraphics[width=0.45\textwidth]{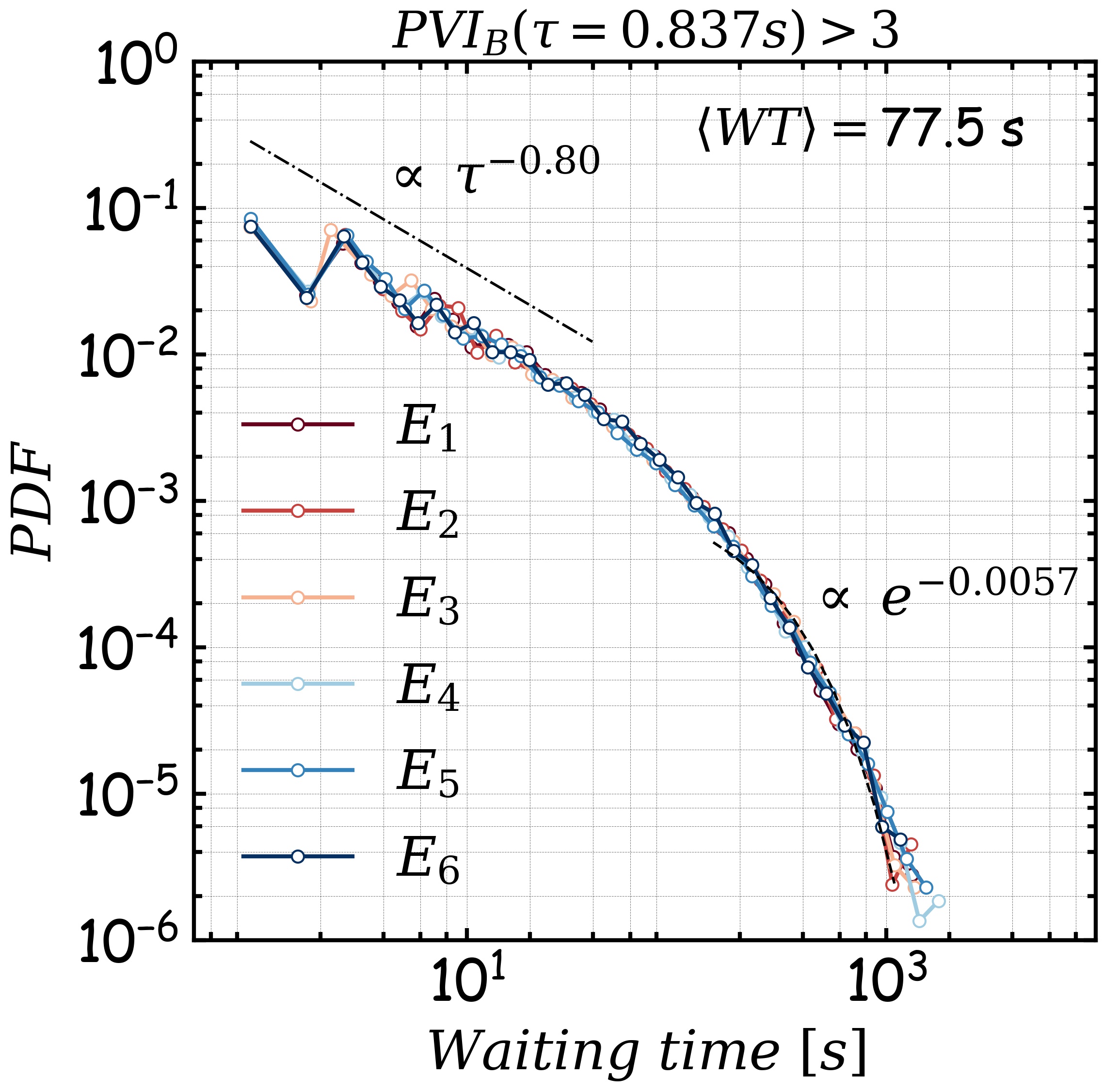}}
\fbox{\includegraphics[width=0.45\textwidth]{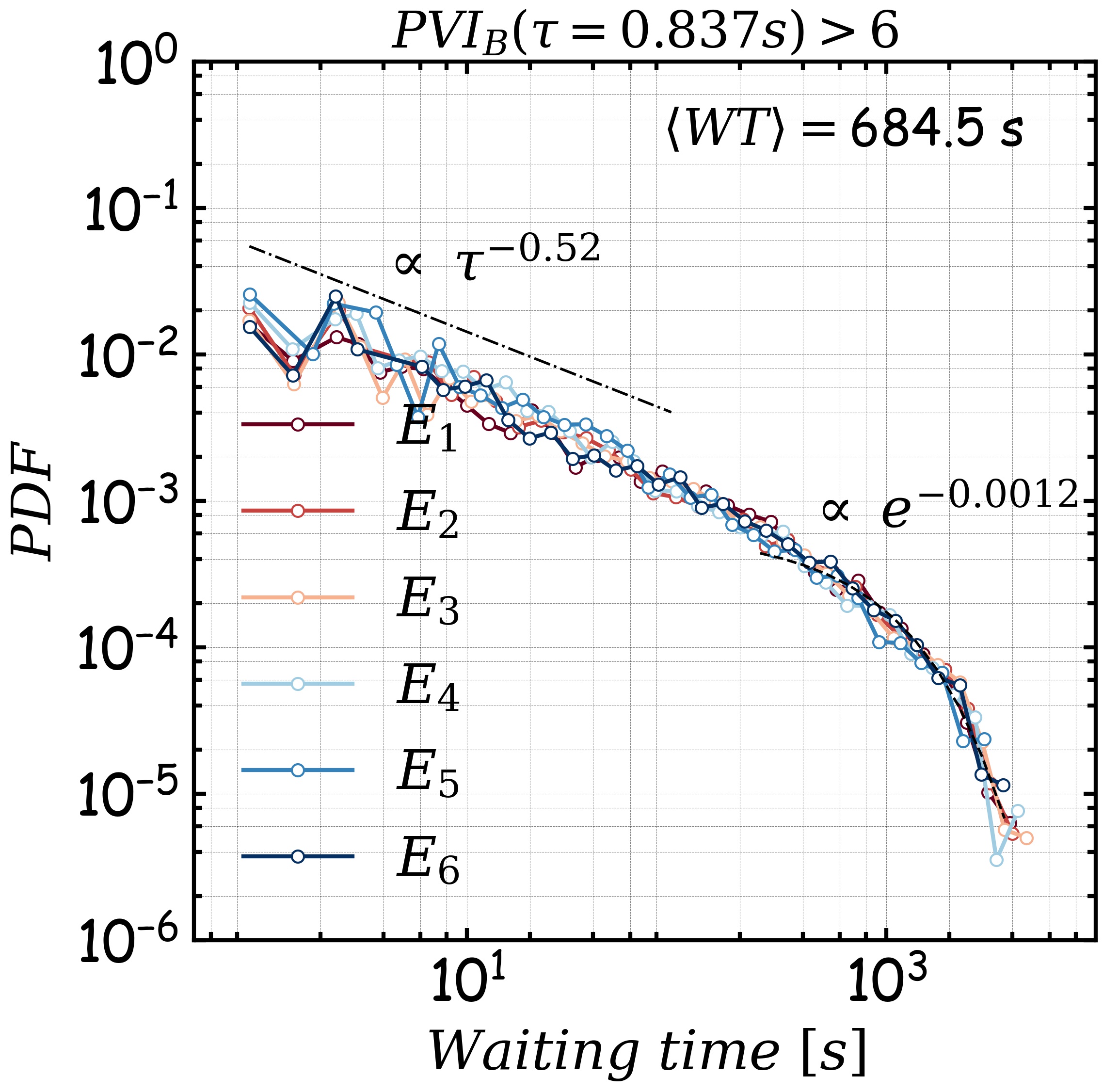}}
\caption{PDFs of WT between (a) $PVI > 3$, (b) $PVI > 6$ events for lag $\tau = 0.837~{\rm seconds}$.}
\label{fig:PVI2}
\end{figure*}

\section{DATA AND ANALYSIS PROCEDURES}\label{sec:data}

We analyze data from first six encounters ($E_{1}-E_{6}$) of PSP from  2018 to 2020, covering heliocentric distances $0.1\lesssim R\lesssim 0.25$~au. We use magnetic field data from the FIELDS fluxgate magnetometers \cite{2016SSRv..204...49B}. To estimate the PVI time-series at time-lags, $\tau \ = \ 0.837~{\rm seconds}$ magnetic field data have been resampled to a  cadence of $0.837~{\rm seconds}$ using linear interpolation. As outlined in Sec \ref{sec:Background}, in order to compute the variance, a moving average over a window that is a multiple of the correlation time is required. The correlation time can be estimated using the e-folding technique by considering the time it takes for the autocorrelation function to drop to $e^{-1}$ of its maximum value \citep{1982JGR....8710347M, 2019ApJ...871...68K}. For encounters $E_{1}-E_{6}$, the correlation time was estimated to  be between  $500\lesssim t\lesssim 2000~{\rm seconds}$. Accordingly, we perform the ensemble averaging over a window of 8 hours for all six encounters.  Several different averaging windows, ranging
$2-12~{\rm hours}$ have been implemented, with qualitatively similar results on our analysis. Plasma data from Solar Probe Cup (SPC), part of the Solar Wind Electron, Alpha and Proton (SWEAP) suite \citep{2016SSRv..204..131K}  have also been analyzed to obtain the bulk velocity and radial proton temperature/thermal speed measurements at $\sim 0.837~{\rm seconds}$ resolution. The radial temperature and bulk velocity time-series have been pre-processed to eliminate spurious spikes and outliers using the Hampel filter \citep{doi:10.1080/01621459.1993.10476339}.
An important effect that should be taken into account when analyzing solar wind particle data is the anisotropy in the parallel ($T_{||}$) and perpendicular ($T_{\perp}$) temperatures with respect to the  background magnetic field
 \citep{Huang_2020, 2011JGRA..116.9105H}. 

\begin{figure*}
\centering
\setlength\fboxsep{0pt}
\setlength\fboxrule{0.0pt}

\includegraphics[width=0.95\textwidth]{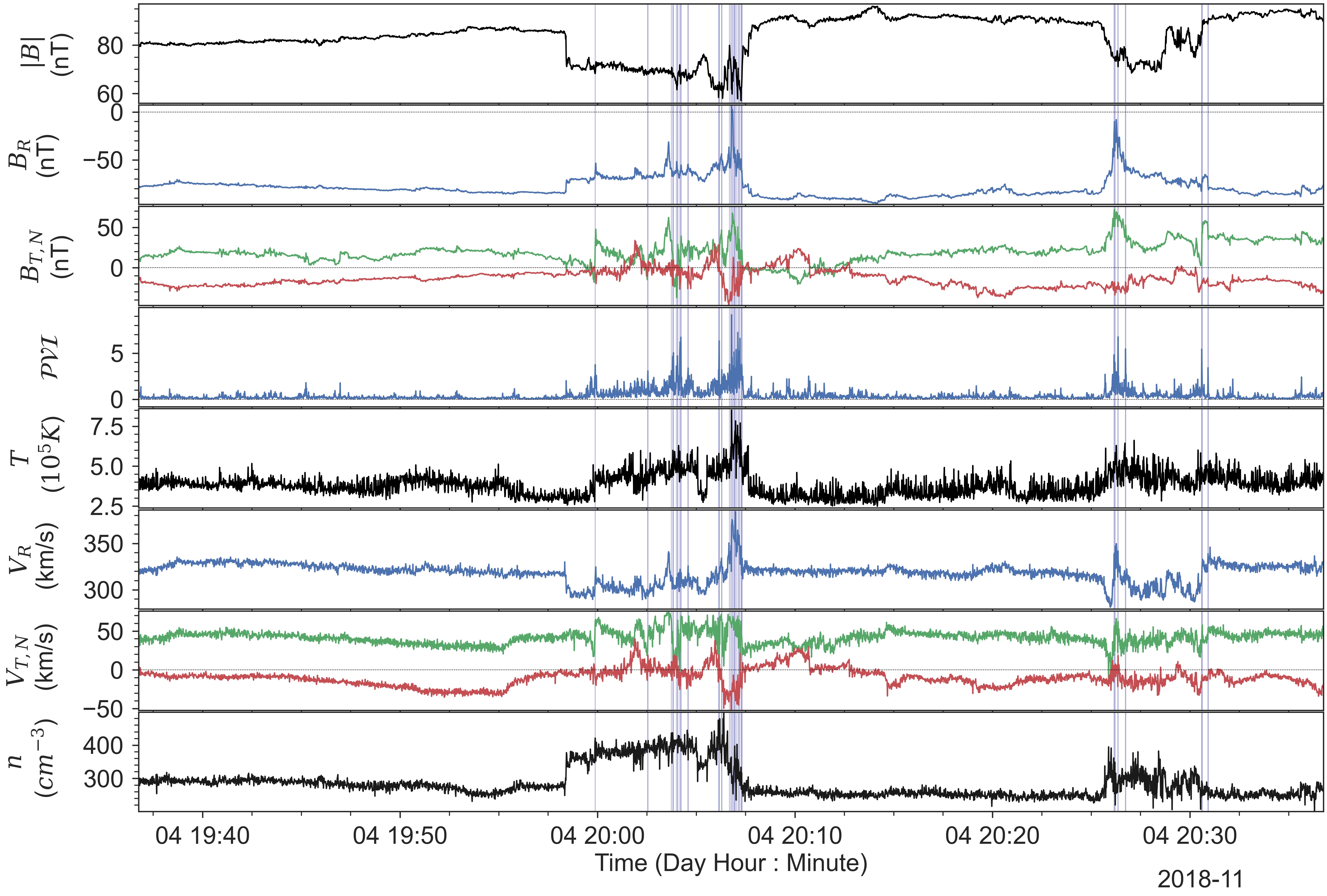}
\caption{An example indicating clustering of intermittent structures associated with increased proton temperature. The shaded areas point to the location of $PVI \geq 3$ events. From top to bottom, the magnitude of the magnetic field $|B|$, the radial component of the magnetic field ($B_{R}$), the tangential and normal components of the magnetic field ($B_{T}$) and ($B_{N}$), in green and red respectively, the PVI time-series for lag,  $\tau = 0.837~{\rm seconds}$  the temperature $T$, the radial component of the proton bulk velocity $V_{R}$, the tangential and normal components of the proton bulk velocity ($V_{T}$), and proton number density ($n$) are shown.}
\label{fig:PVI3}
\end{figure*}


SPC was designed to sample the radial temperature. Consequently, to ensure that the temperature
we were measuring was indeed radial temperature all intervals for which the angle between $B_{r}\hat{r}$ and $\textbf{B}$ was greater than thirty degrees, $\theta_{B_{r}\hat{r},\textbf{B}} \geq 30$ have been discarded. Finally, intervals for which 
the solar wind was outside of the field of view of SPC, have also been manually discarded.

\section{Results}\label{sec:results}

\subsection{Statistical properties of intermittent structures.}\label{subsec:Stat_properties}
One of the major questions one has to address when studying intermittency is the nature of the physical processes that initiate the coherent structure production at the origin. Accordingly, in order to gain insight into the statistics of intermittent magnetic structures in the solar wind, we follow the process described in Sec. \ref{sec:data} to estimate the PVI time-series for time-lag, ${\tau} \ = \ 0.837~{\rm seconds}$. In Fig. \ref{fig:PVI1}a we show the Probability Density Functions (PDFs) of PVI values. The most probable value is close to $\sim \ 0.3$,  indicating that the majority of the detected events can be characterized as non-intermittent. In Fig. \ref{fig:PVI1}b, the fraction of the entire data-set occupied by PVI events exceeding the threshold $PVI > \theta$, is shown  for $E_{1}-E_{6}$. For reasons that will become obvious in Sec \ref{subsec:heating}, the fraction of events characterized by a PVI index greater than unity $f_{PVI \geq 1 }$ is also shown. The fraction is consistent for $E_{1}-E_{4}$ attaining a value of $f_{PVI \geq 1 } \approx 19.1 \% $, but decreases to  $f_{PVI \geq 1 } \approx 18 \% $  for $E_{5},E_{6}$. Beyond this point, the fraction of the dataset occupied by coherent structures characterized by higher values of PVI decreases rapidly. 

\begin{figure*}
\centering
\setlength\fboxsep{0pt}
\setlength\fboxrule{0.0pt}
\fbox{\includegraphics[width=0.95\textwidth]{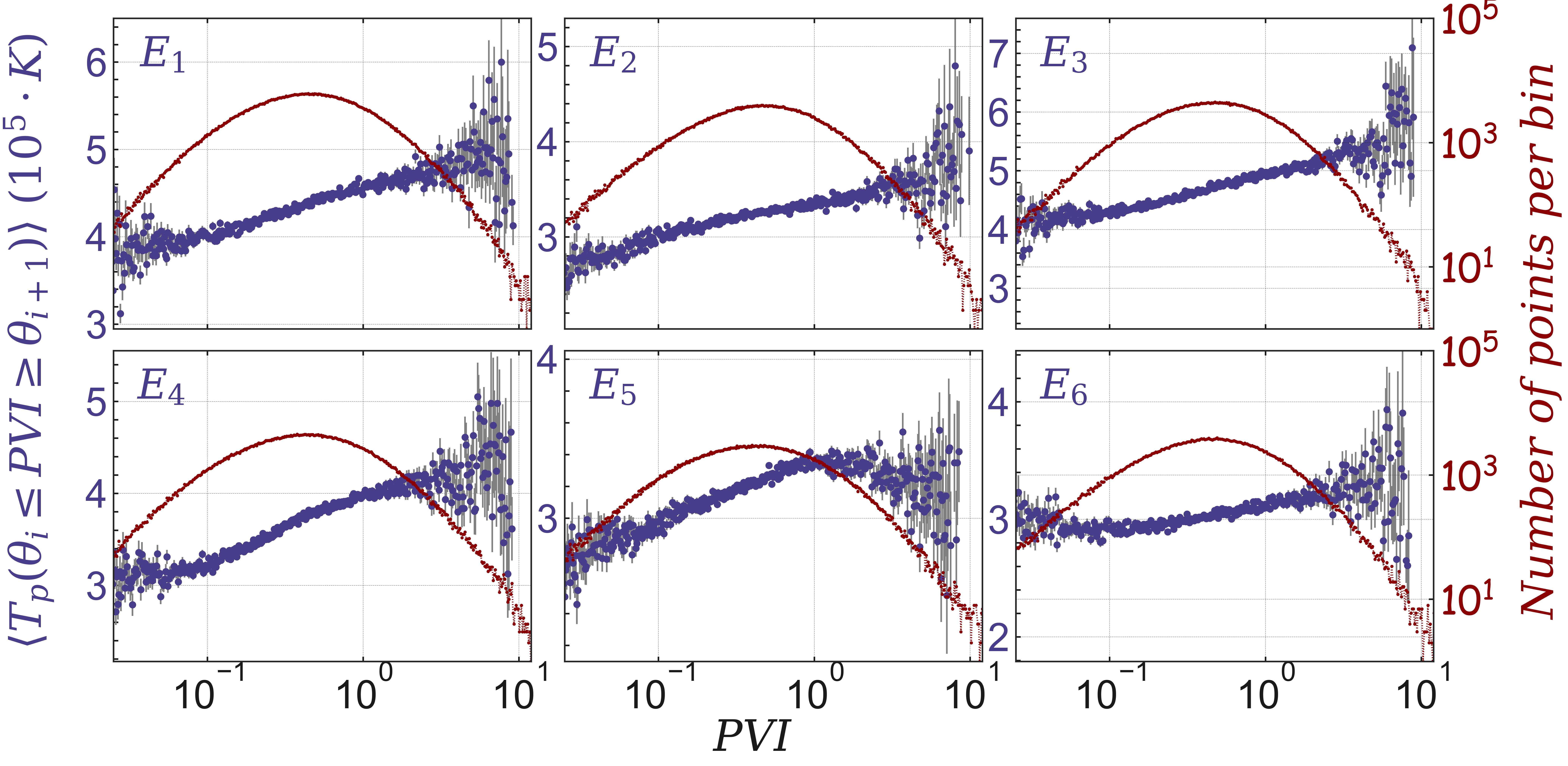}}
\caption{Binned average of radial proton temperature plotted against PVI (blue) along with the number of points per bin (red) for the first six encounters of PSP. Notice a rough upward trend in mean temperature at higher PVI bins. }
\label{fig:heating1}
\end{figure*}


\par Another method that can provide an insight into the statistics of the solar wind coherent structures is the Waiting Time (WT) distribution analysis. In the case of the PVI time-series, we define the waiting time as the time passed between the end and the start of two subsequent events for which the value of PVI stays above a threshold $\theta$. Note, that PVI events have a finite duration. Therefore, subsequent times for which the PVI time-series stay above the threshold are considered as part of the same event. Also note that in order to maintain an adequate sample size,  we have imposed a restriction on the minimum counts per bin. Consequently, bins with fewer than ten counts have been discarded. The waiting time interval is itself a new random variable, the distribution of which, is independent of the original random variable's distribution. A simple inspection of the distribution shape can then reveal whether or not the underlying mechanism can be classified as a random Poissonian-type of process, or it possesses “memory” indicating strong correlation and clustering. In the first case, the distribution is better described by an exponential, while in the latter distribution, scales like a power-law \citep{PhysRevE.80.046401, Greco10}. Figs. \ref{fig:PVI2}a,b, show the PDF's of WT between intermittent PVI events with lag $\tau \ = 0.837~{\rm seconds}$ and threshold ${\theta}_{1} =  3$, ${\theta}_{2} = 6$, respectively, for the first six encounters. At lower WT's, the best fit analysis indicates that WT distributions are better described by a power-law. The index of the power-law fit attains values in the range $a \ \in [-0.8, \ -0.52]$, progressively getting softer as the threshold value, $\theta$, increases. In contrast, for events further apart in time, the distribution is better described by an exponential. The change between power-law and exponential in the distribution can be interpreted as the breaking-point between $intercluster$ and $intracluster$ waiting times \cite{Greco10}. This change seems to coincide with an ill-posed, due to the power-law nature of the distribution, mean value of WT,  $\langle WT \rangle$ \cite{Chhiber_2020}. The WT distribution analysis was repeated by estimating the PVI time-series using a different time-lag, $\tau \ = \ 8.37, \  83.7~{\rm seconds}$, still sampling though, overtimes with $0.837~{\rm seconds}$ cadence. The resulting waiting time distributions (not shown here) are once again remarkably similar between all six encounters and thus resemble the ones reported in \cite{Chhiber_2020} for $E_{1}$. This provides a strong indication that high PVI valued coherent structures, are not evenly distributed within the solar-wind, but rather tend to be strongly correlated and form clusters, creating alternating regions of very low and very high magnetic field activity respectively (see also  \citep{2020ApJS..246...39D, Chhiber_2020, 2021ApJ...923..174B,  2021ApJ...921...65Y}). One such example, out of the thousands of clustering events we were able to recover is presented in Fig. \ref{fig:PVI3}. To emphasize the clustering of coherent structures, blue, vertical lines have been added to indicate the location of events characterized by a magnetic PVI index, $PVI \geq 3$. Notice the elevated proton temperature at regions where coherent structures abound, an effect that is discussed in detail in Sec. \ref{subsec:heating}.

\begin{figure*}
\centering
\setlength\fboxsep{0pt}
\setlength\fboxrule{0.0pt}
\fbox{\includegraphics[width=1\textwidth]{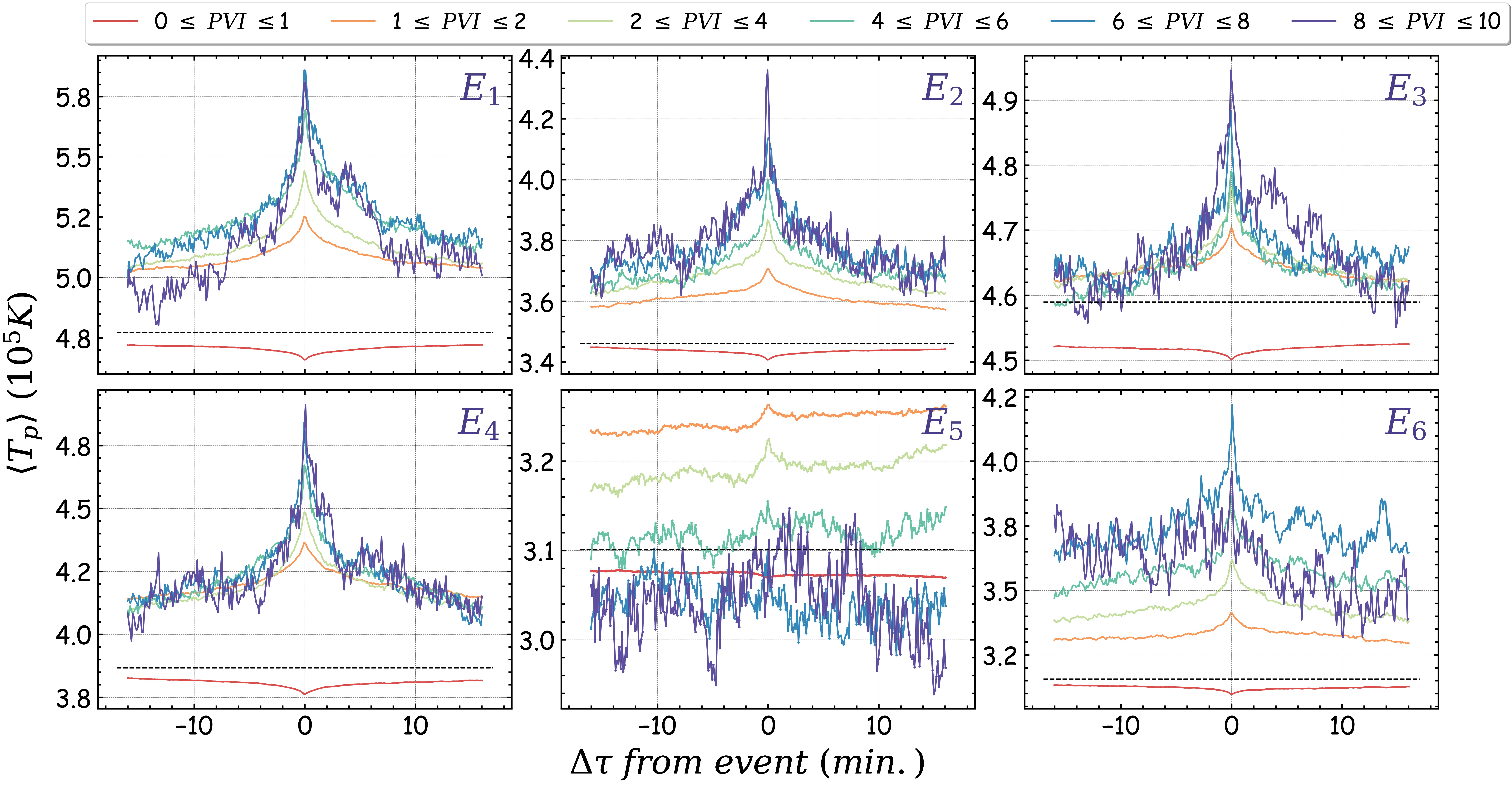}}
\caption{Average $T_{p}$ conditioned on the temporal separation from PVI events that exceed a PVI threshold. The black dashed line indicates the mean proton temperature ($\tilde{T}_{p}$) for each encounter. }
\label{fig:heating2}
\end{figure*}


\begin{figure}
\centering

\includegraphics[width=0.42\textwidth]{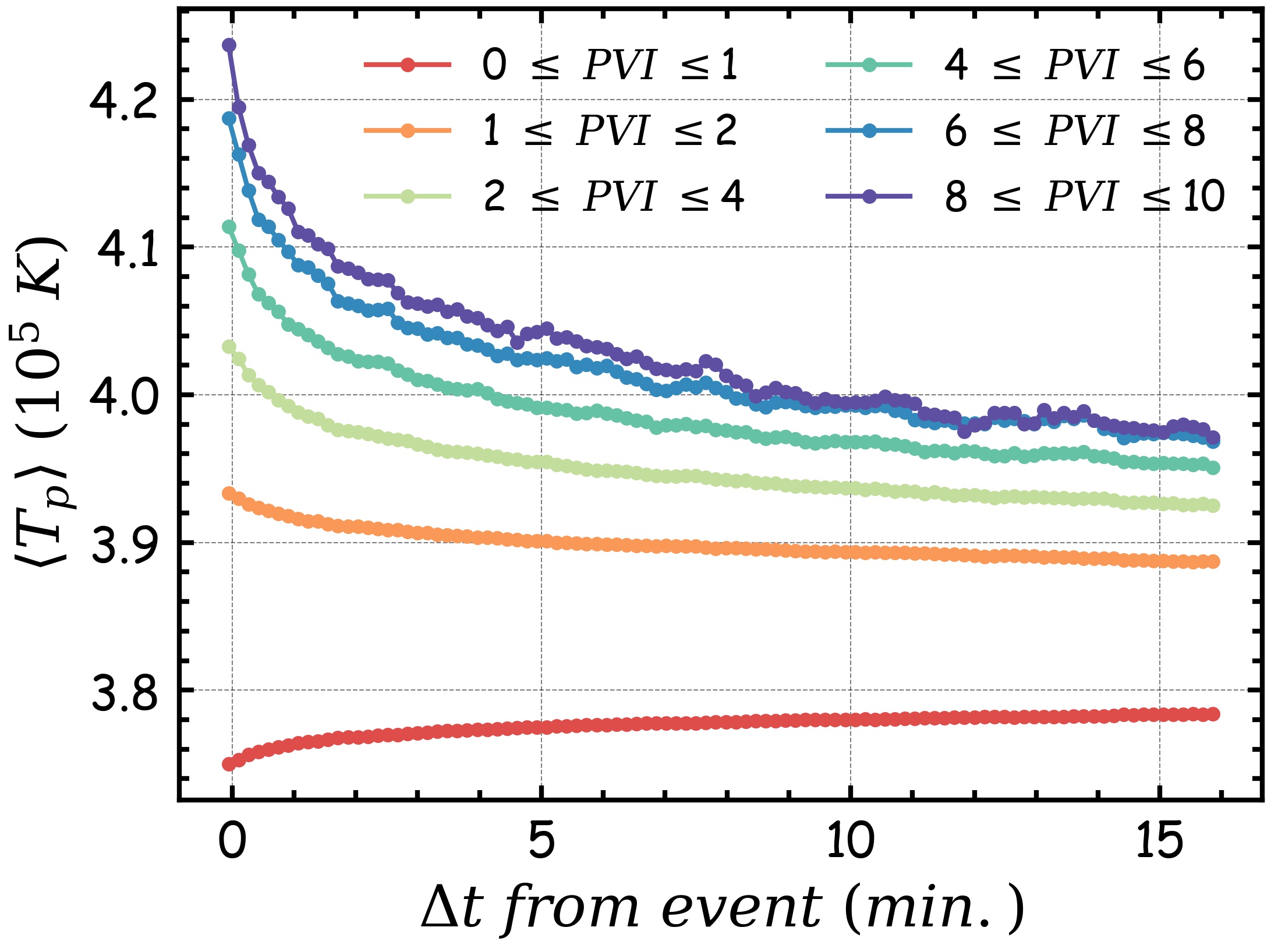}
\caption{Superposed average of $T_{p}$ conditioned on the temporal separation from PVI events that exceed a PVI threshold  for $E_{1}-E_{6}$}
\label{fig:heating3}
\end{figure}


\subsection{Intermittent heating of the young solar wind.}\label{subsec:heating}

The purpose of this section is to investigate the idea that local coherent structures contribute to the heating of the solar wind. For this reason, we follow the method described in Sec. \ref{sec:data} and carry out several diagnostics on the correlation between the PVI and proton temperature ($T_{p}$) time-series. As a first step, we use binned statistics to interpret $T_{p}$ as a function of PVI. In Fig. \ref{fig:heating1}, we use 600 PVI bins, to present the average proton temperature per bin (i.e.,  $\langle T_{p}(\theta_{i} \leq PVI \leq \theta_{i+1}) \rangle$, where $\theta_{i}$ is the PVI threshold) plotted against the center of the bin.  Uncertainty bars are also shown as vertical lines indicating the standard error of the sample \citep{10.2307/2682923}. The uncertainty is thus estimated as $\sigma_{i}/ \sqrt{n}$, where $\sigma_{i}$ is the standard deviation of the samples inside the bin. The number of samples included in each PVI bin is also shown as a dotted red line. For all encounters, a statistically significant, positive correlation between $T_{p}$ and PVI can be observed for $PVI \leq 3$. Beyond this point, the trend is eroded for $E_{5}$. 
For the remaining encounters, the trend extends, with good statistical significance to $PVI \approx 6$. For greater values of PVI, $6 \leq PVI \leq 10$, though on average the temperature increases with increasing PVI, there is also a high degree of variability that could most probably be attributed to the low number of recorded events, as indicated by the red dotted line. For most orbits, the local temperature peaks at $PVI \sim 10$. Between the lowest and highest PVI values, the difference in mean temperature 
 is of the order of $\sim 10^{5}K$, but, can range as high as $\sim 2 \cdot 10^{5}K$ (e.g., $E_{1}, \ E_{2}, \ and \ E_{4}$). Note, however, the high degree of variability in $\tilde{T_{p}}$ for bins in the $PVI \geq 6$ range. Although bins of relatively high average temperature are the rule, several bins of low temperature, comparable to one of the bins of $PVI \approx 2$,  can be observed. \par

\begin{figure*}
\centering
\setlength\fboxsep{0pt}
\setlength\fboxrule{0.0pt}

\fbox{\includegraphics[width=7.3in]{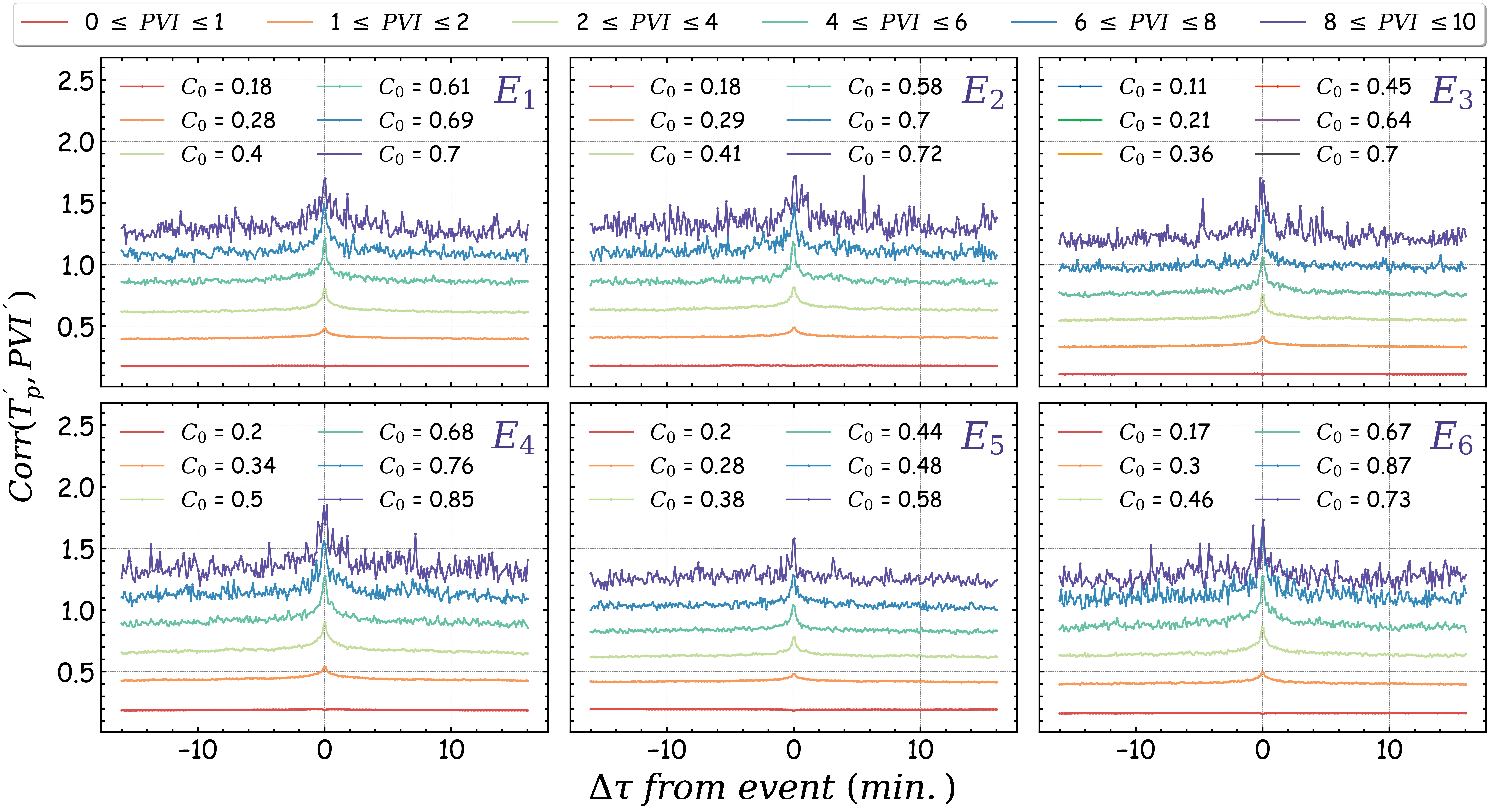}}

\caption{ Lagged cross correlation between $ T_{p}^{'}$ and $ PVI^{'}$ for different thresholds in the PVI timeseries. The correlation coefficient at zero lag is also shown. Note that the correlation functions have been shifted in the vertical direction
for clarity.}
\label{fig:heating4}
\end{figure*}


To further elucidate the relationship between $T_{p}$, and magnetic field discontinuities, we estimate averages of $T_{p}$ constrained on the temporal separation from PVI events that lay in a given PVI bin. This can be expressed as \citep{PhysRevLett.108.261102, Tessein_2013, article}:

\begin{multline}
     \tilde{T}_{p}(\Delta t, \ \theta_{i}, \ \theta_{i+1})  \ = \\ \langle T_{p}(t_{PVI} \ + \Delta t)| \theta_{i} \leq PVI \leq \theta_{i+1}\rangle.
\end{multline}

Here, $\Delta t$ is the temporal lag relative to the location of the main PVI event taking place at time $t_{PVI}$, and $\theta \ = \ [0, \ 1, \ 2, \ 4,  \ 6,  \ 8, \ 10]$.  The conditional average of temperature at different lags is presented in Fig. \ref{fig:heating2} for six different PVI bins. The dip in the mean temperature at lag equal to $t \ = 0 s$, for $PVI \leq 1$ (red line), suggests that no significant heating of the solar wind occurs at times where the magnetic field is relatively smooth. This observation is reinforced by the fact, that in such regions, the plasma temperature is lower than the mean temperature of the respective encounter, indicated by the black dashed line. On the other hand, by increasing the threshold value $\theta$, we can make several observations regarding the nature of intermittent dissipation in the solar wind. For encounters $E_{1}-E_{4}, \& \ E_{6}$, and $\theta \geq 1$, we can observe a global maximum in mean $T_{p}$ close to zero lag, indicating a higher probability of increased proton temperature in the vicinity of coherent structures. The peak is then followed by a gradual decrease in mean $T_{p}$ as we move further away from the main discontinuity. The fact that $T_{p}$ stays elevated near the main event can, most probably, be attributed to the clustering of coherent structures noted ins Sec. \ref{subsec:Stat_properties}. The rate of decrease is distinct for each bin, with the steepest gradients in $T_{p}$ observed around the sharpest discontinuities. This is better illustrated in Fig. \ref{fig:heating3}, where a superposed average of $E_{1}-E_{6}$ is presented. \par


\begin{figure*}
\centering
\setlength\fboxsep{0pt}
\setlength\fboxrule{0.0pt}

\fbox{\includegraphics[width=7.3in]{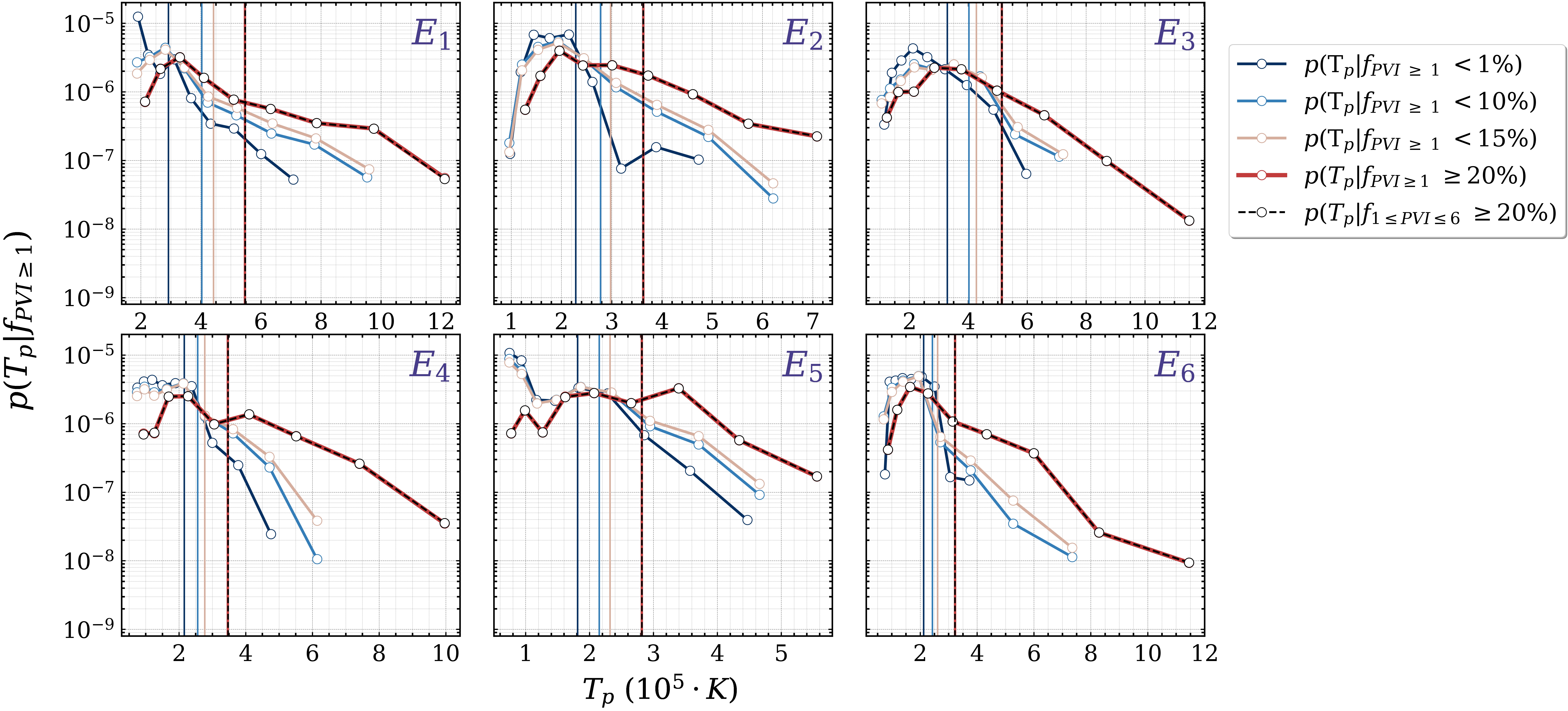}}

\caption{ PDF's of $T_{p}$ conditioned on the percentage fraction of coherent structures with $PVI \geq 1$, $f_{PVI \leq 1}$ for the first six encounters. Vertical lines indicate the median temperature for each of the PDFs. As the percentage of high PVI valued events increases, we can see a significant increase in the probability density for higher proton temperatures.   Note that PDFs don't change when we remove  $PVI \geq 6$ events (black, dashed line).}
\label{fig:heating5}
\end{figure*}


\begin{figure}
\centering

\includegraphics[width=0.45\textwidth]{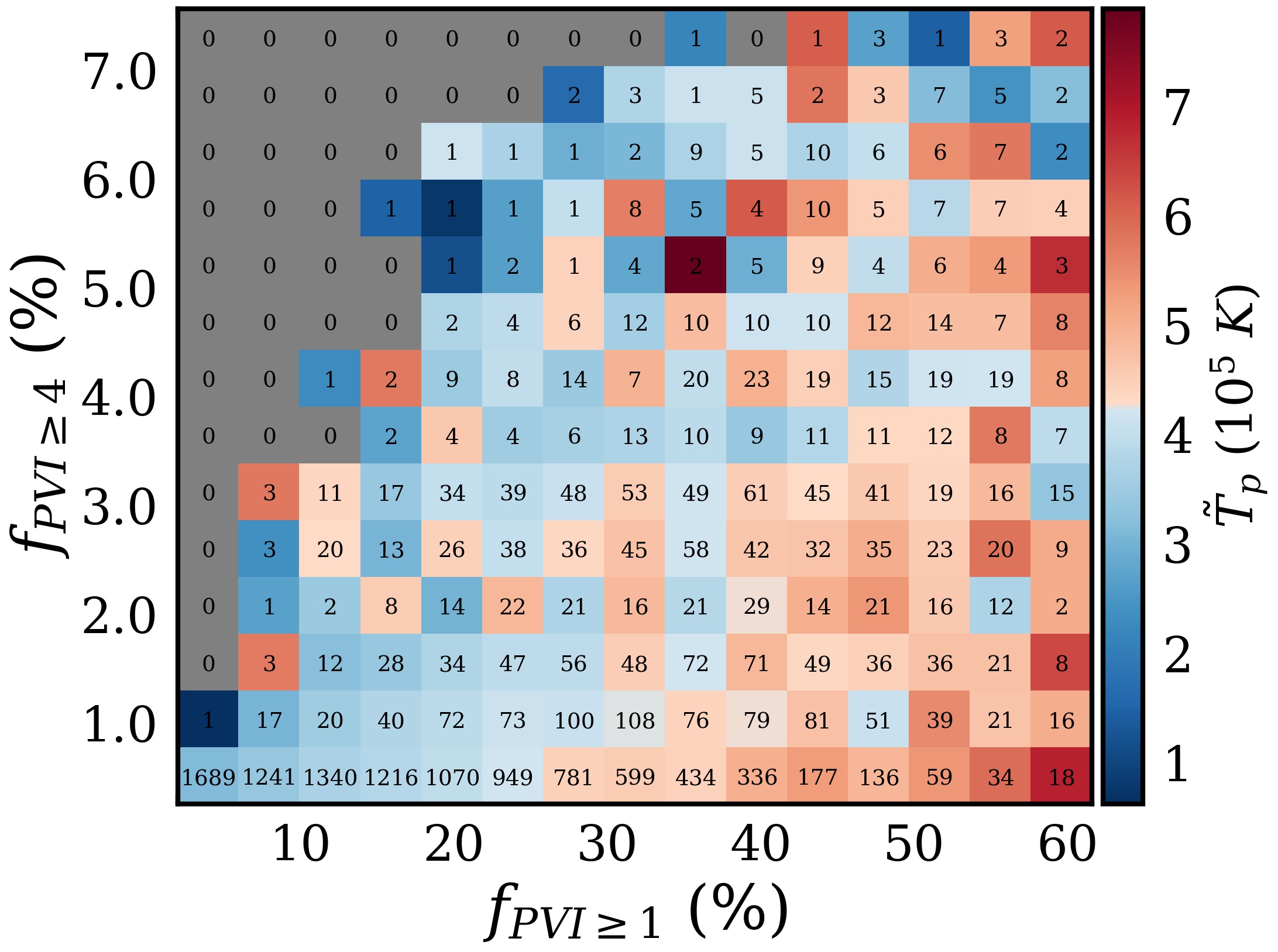}

\includegraphics[width=0.45\textwidth]{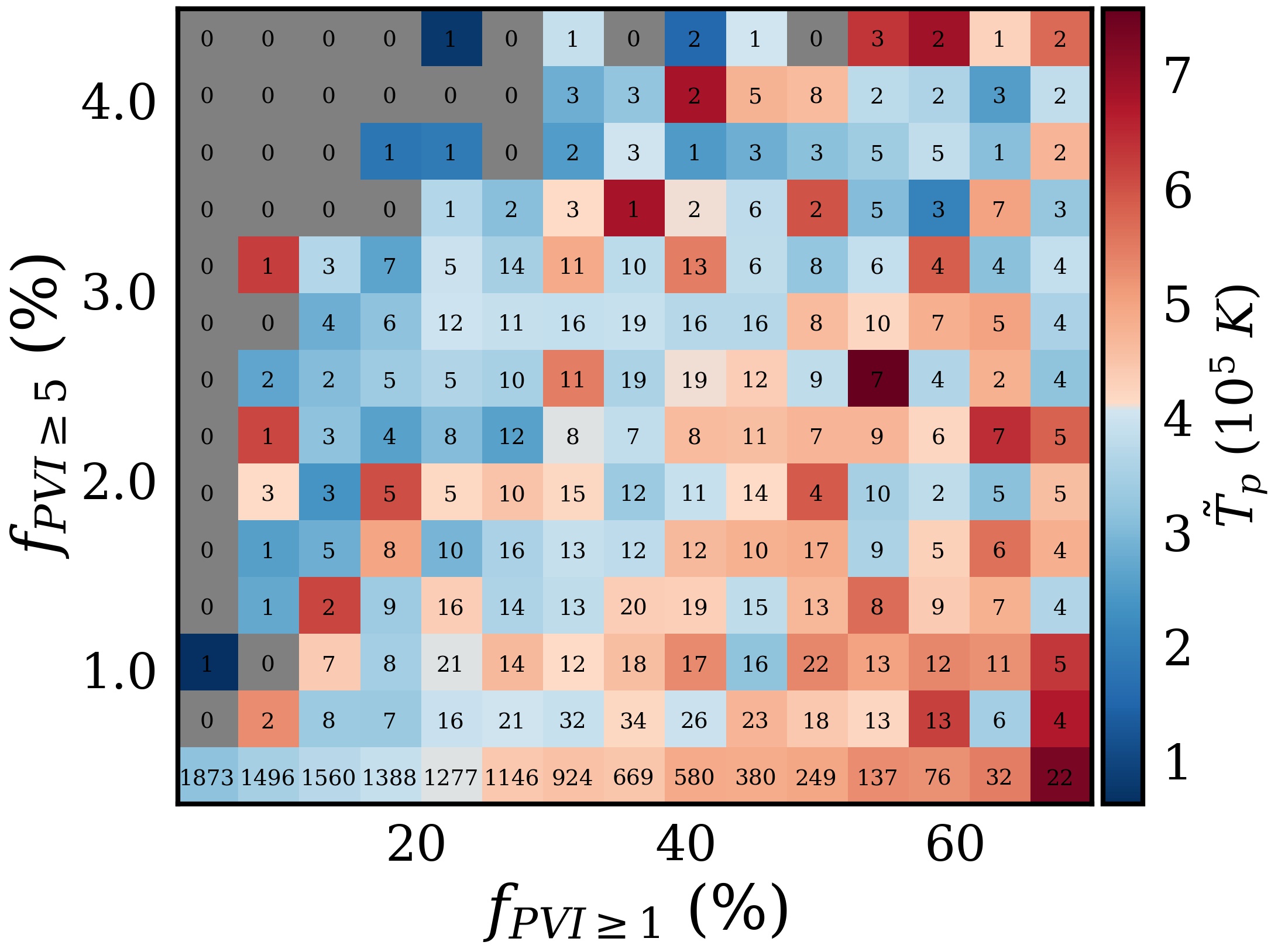}

\includegraphics[width=0.45\textwidth]{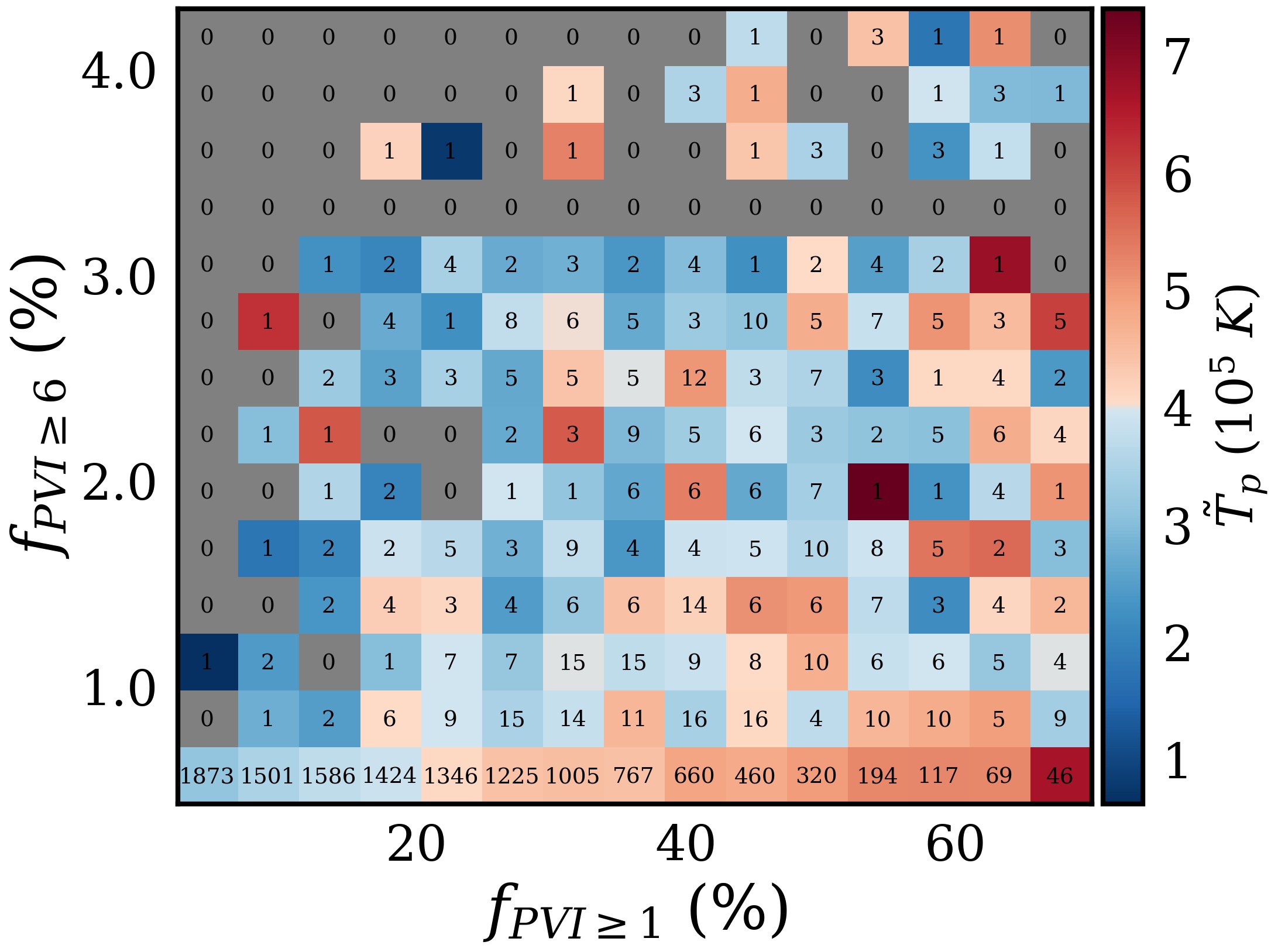}
\caption{  Fraction of coherent structures of PVI index greater than $\theta$, $f_{PVI \geq \theta}$, where (a) $\theta$ = 4,  (b) $ \theta$  = 5, (c) $\theta $= 6 as a function of the fraction of coherent structures of PVI index greater than one $f_{PVI \geq 1}$  and mean proton temperature of the 5-minute interval $\Tilde{T_{p}}$. The data points were binned according to $f_{PVI \geq \theta}$ and $f_{PVI \geq 1}$, and the mean value inside each bin was calculated, which is reflected in the color. The numbers in the plot indicate the number of 5-minute intervals inside each bin. Bins with 0 counts are shown in dark grey.}
\label{fig:heating6}
\end{figure}


On the contrary, 
$E_{5}$ exhibits a  distinct behavior. The mean proton temperature is  $T_{p} \ \sim 3 \cdot 10^{5}K$  marking the lowest observed proton temperature between all encounters, and the average $T_{p}$ peaks at lower thresholds of PVI. These results motivated us to perform a more thorough investigation of $E_{5}$. 
A possible explanation for the observed discrepancy during E5 is the crossing of PSP through the Heliospheric Current Sheet ($HCS$). Several statistical studies using superposed epoch analysis suggest that the HCS is characterized by a local minimum in proton temperature, and a local maximum in the proton density \citep{1981JGR....86.4565B, 2009JGRA..114.4103S, 2021ApJ...920...39L, shi2022influence}. As a result of the extended intervals characterized by an increased proton density in the vicinity of the HCS, transient mechanisms such as adiabatic proton heating can obscure the effects of non-adiabatic, intermittent heating. Moreover, recent observations \citep{article_chen}, support a poorly developed turbulence in the proximity of the HCS, indicated by the relatively lower amplitude of the turbulent fluctuations. These results are corroborated by the fact that $E_{5}$, is characterized by a relatively low coherent structure density. More specifically, the precise value for the fraction of the data-set occupied by events of $PVI \geq 1$, was estimated at $f_{PVI \geq 1 } \approx 18 \%$, namely, $6.1\% $ lower than the mean value of  $\langle f_{PVI \geq 1 } \rangle = 19.1 \%$ for the rest of the encounters. The poorly developed turbulence in the vicinity of the HCS could act as a contributing factor to the ineffective heating of the protons in that region. Note, however that other features of the solar wind, including the Alfvénic nature of the fluctuations (e.g. through cross-helicity) \citep{2019MNRAS.482.1706S}; solar wind speed \citep{1970ApJ...159..659B,1986JGR....91.1701L,  2021A&A...650A..21S}; plasma beta \citep{2017ApJ...850L..11V, 2020ApJS..246...70H}; the presence of waves \citep{2012ApJ...753L..19H, 2015ApJ...813L..30H, 2021ApJ...919L...2M} have been shown to be closely correlated with proton temperature. Taking all the aforementioned effects into account, we can understand that the  intermittent heating in the solar wind, could be obscured by other transient heating mechanisms that are strongly dependent on the nature of the solar wind under study, and could explain the discrepancy for $E_{5}$. \par 
The results presented up to this point provide strong evidence for the intermittent heating of the solar wind. However, for a more complete understanding, it would be convenient to quantify the relative contribution of different types of coherent structures, as indicated by the PVI index,  to the solar wind's internal energy. To do this, a careful analysis that allows for clustering effects, as well as transient effects (e.g. HCS crossing) to be subtracted is required. One such analysis involves the study of lagged cross-correlation between $T_{p}$ and the PVI time-series, as it offers the chance to infer whether or not the two quantities are changing at the same time. Here, the lag refers to how far the series is offset and its sign determines which series is shifted, in this case, PVI being the lagged quantity. The goal is to correlate the fluctuating parts of PVI and $T_{p}$. We thus have to subtract the mean 
\begin{equation}
    PVI^{'} \ = \ PVI(t) \ - \ \langle PVI \rangle,
\end{equation}
\begin{equation}
    T_{p}^{'} \ = \ T_{p}(t) \ - \ \langle T_{p} \rangle,
\end{equation}

where $\langle ... \rangle$ denotes the time average over the entire data set. 
We can therefore estimate the lagged cross-correlation as
\begin{equation}
    Corr( T_{p}^{'},  PVI^{'}) \ = \ \frac{\langle  PVI^{'}(t  \ + \ \tau )  \cdot  T_{p}^{'}(t) \rangle}{\sqrt{\sigma_{PVI} \cdot \sigma_{T_{p}}}},
\end{equation}  \label{eq:Correlation} 
where $\sigma$ is the variance of the subscripted quantity. In Fig. \ref{fig:heating4}, the lagged cross-correlation between PVI and temperature, for different $PVI$ bins is presented. Note that each curve has been shifted vertically
by $0.2 \cdot i$, where,  $i \ = \ 0,1,..,5$ for better visualization. \par
For $PVI \leq 1$,  we can observe a minimum of the correlation coefficient at zero lag providing an additional confirmation that no temperature enhancements occur in such regions. On the other hand, for higher thresholds, there is a trend, with the correlation coefficient attaining lower values as we move further away from the discontinuity. In general, the maximum value of the coefficient tends to peak at higher values with the increase of PVI. In many cases, though, the three highest PVI bins peak at comparable values. For $E_{5}$, even though the correlation coefficient stays relatively low, peaks still emerge at the location of the principal discontinuity, at zero lag. We can thus understand that the strongest discontinuities in the magnetic field are related to the most abrupt changes in temperature, and thus play a prime role in determining the local plasma dynamics.
Finally, note the emergence of local peaks at some distance from the main discontinuity, becoming more pronounced as we move to higher PVI thresholds. We can attribute this observation to the combination of two factors. The first factor is the relatively low frequency of high PVI valued events. Due to the scarcity of such events, around which we perform the conditional averaging, the number of samples gets progressively lower as the PVI threshold increases. For example, for PVI index, $PVI \geq 8$, the number of recorded events for $E_{1}-E_{6}$ ranges between 140 to 277. Consequently, even though there are enough events at the largest PVI bins to safely determine the mean and variance in the correlation coefficient, single strong events of PVI index, $PVI \geq 6$, shown to be highly correlated to local temperature changes, located at a given distance from the main discontinuity might be controlling the correlation. The second factor is the clustering of coherent structures reported in Sec.\ref{subsec:Stat_properties}. For the clustering effects, it would be important to also take into account events of PVI  index, $PVI \geq 6$, with the number of recorded events for $E_{1}-E_{6}$ ranging between 456 to 820. As shown in Fig. \ref{fig:PVI2}b, the WT distribution for $PVI \geq 6$ is described by a power-law at lower waiting times indicating that such events tend to be highly correlated and form clusters. Consequently, considering clustering effects, we can understand that intense events taking place close to the main event can strongly affect the average value of the mean $T_{p}$ at certain lags, resulting in the local peaks observed in Fig. \ref{fig:heating4}. 


Finally, we move on to study the macroscopic effects of intermittent heating. For this reason, we divide the data into $5 \ minute$ sub-intervals, and for each sub-interval, the fraction of the data that correspond to events of PVI index greater than unity, as well as, the mean temperature of the sub-interval are estimated. This allows us to estimate the PDFs of the mean proton temperature, $T_{p}$, conditioned on the percentage fraction of $PVI \geq 1$ events within the sub-interval, $p(T_{p}|f_{PVI \geq 1} \leq \theta \%)$, where, $\theta \ = \ 1, \ 10, \ 15, \ 20$. As shown Fig. \ref{fig:heating5},   the probability density for a higher mean temperature is increased, as the fraction of coherent structures in the interval increases. Additionally, the median value of proton temperature within each bin is estimated, shown as a vertical line of the same color to the respective PDF. It is clear from Fig. \ref{fig:heating5} that for denser in coherent structure intervals, the median of $T_{p}$ moves to higher values. Notice that even when we exclude all events characterized by $PVI \geq 6$, usually associated with reconnection exhausts, from our analysis, both the PDFs and the median $T_{p}$ practically remain intact.  Taking into account the steep temperature gradients associated with events of high PVI index,  shown in Fig. \ref{fig:heating5}, the results of Fig. \ref{fig:heating5} may at first seem contradictory. However, the negligible contribution of events characterized by a high PVI index can be explained in terms of the relatively low frequency of occurrence of such events, as shown in Fig. \ref{fig:PVI1}b. This indicates, that as a result of their scarcity, coherent structures characterized by a PVI index that exceeds a given threshold, dissipate a negligible amount to the internal energy of the solar wind. To better illustrate the relative contribution, of coherent structures at a given threshold $\theta$,  $PVI \geq \theta $, we present in Fig. \ref{fig:heating6}a,b,c the fraction of coherent structures of PVI index greater than $\theta = 4, \ 5 , \ 6$  respectively as a function of the fraction of coherent structures of PVI index greater than unity, $f_{PVI \geq 1}$ and mean proton temperature of the 5-minute interval $\Tilde{T_{p}}$. In these figures, the color indicates the mean proton temperature $T_{p}$ of the 5-minute interval, while, the numbers in the plot indicate the number of 5-minute intervals inside each bin. As shown in Fig. \ref{fig:heating6}c, the vast majority of the intervals are characterized by a very low number density, $f_{PVI \geq 6}$, as well as, a clear increase in $T_{p}$ with increasing $f_{PVI \geq 1}$. On the other hand, for $\theta \geq 6$ even though intervals with the highest observed temperature are characterized by a relatively high number of $f_{PVI \geq 6}$, not a clear trend with increasing $f_{PVI \geq 6}$ is observed. At the same time, such intervals constitute only a small, almost negligible fraction of the entire dataset. On the contrary, as shown Fig. \ref{fig:heating6}a,b the number of intervals with a high coherent structure number density and elevated $\tilde{T}_{p}$ steadily increases as we lower the threshold to $\theta = 4,5$, respectively. \par

\begin{figure} \label{fig:heating7}
\centering

\includegraphics[width=0.45\textwidth]{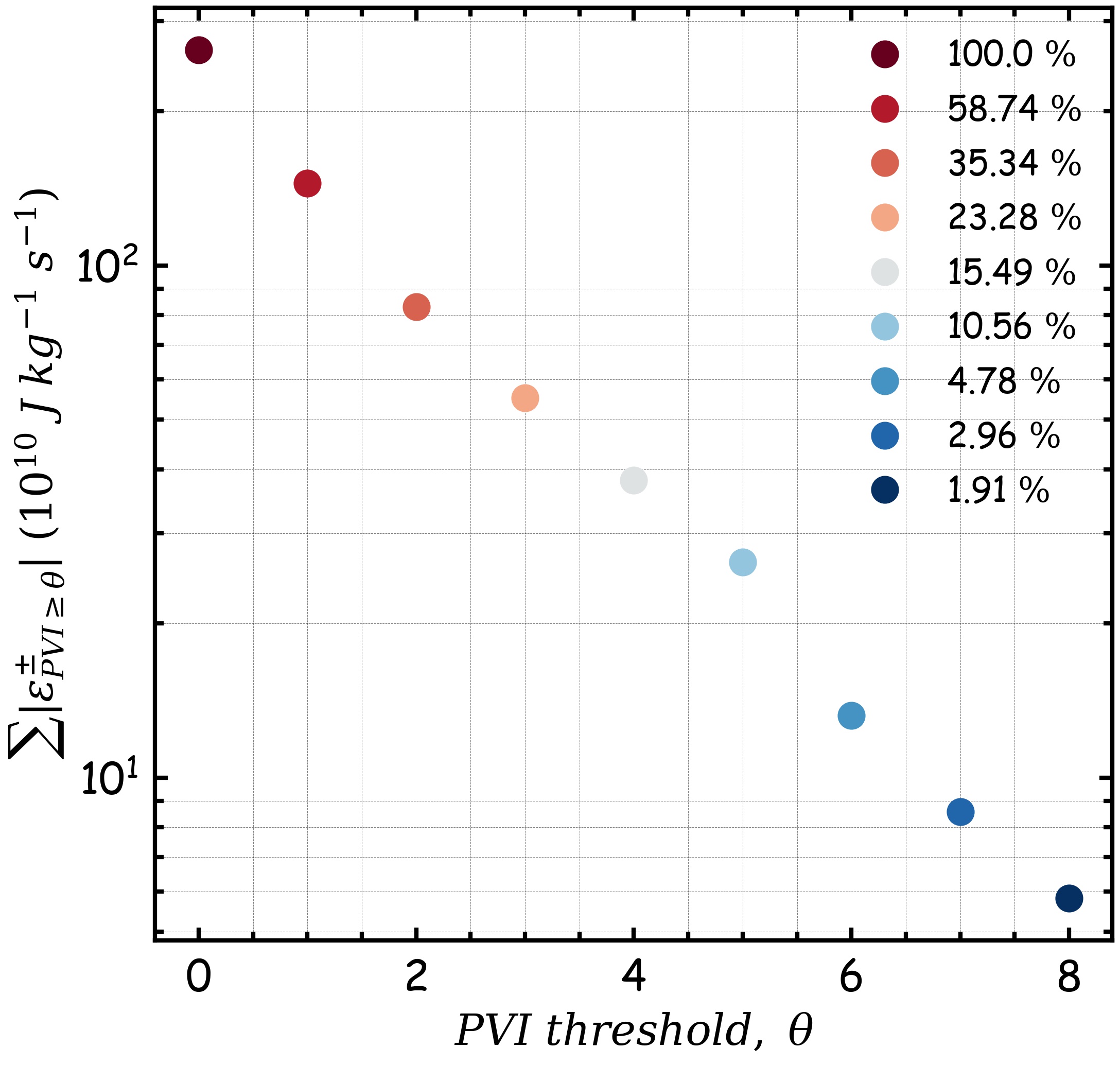}

\caption{ Sum of absolute values of LET, associated with PVI events of a given PVI threshold, $\sum |\epsilon^{\pm}_{PVI \geq \theta}|$. The legend indicates the estimates of the percentage fraction of total LET for coherent structures at a PVI threshold, $ 100 \cdot \frac{\sum |\epsilon^{\pm}_{PVI \geq \theta}|}{\sum |\epsilon^{\pm}_{PVI \geq 0}|}$.} 
\end{figure}

Finally, to further corroborate the previous results, and quantify the relative contribution of coherent structures we take advantage of the LET method (Sec. \ref{sec:Background.2}). In Fig. \ref{fig:heating7}, we estimate the sum of local energy transfer associated with coherent structures characterized by a PVI index that exceeds a given threshold. Note, that LET is a signed quantity. Consequently, for the purposes of this study, the absolute value of LET is considered \citep{article}. In agreement with \citep{PhysRevLett.108.261102}, the strongest PVI events, $PVI \geq 5$, contribute $\sim 10.5 \ \%$ of the internal energy. However, as the PVI threshold is increased to $PVI \geq 6$, the contribution of coherent structures decreases to $\sim 4.8 \ \%$. These results indicate that in addition to the amount of energy dissipated per identified event, future studies should also consider the frequency of occurrence of such events when considering intermittent dissipation in the solar wind. Consequently, even though, intense events of $PVI \geq 6$, usually linked to reconnection exhausts, are the ones that determine the local plasma dynamics, due to the scarcity of such high PVI valued coherent structures, it is rather the accumulative effect of many small scale dissipation events that determine the global temperature of the solar wind. More specifically, our results, suggest that in the near sun solar wind environment the number density of coherent structures characterized by a PVI index in the range $1\lesssim PVI \lesssim 6$, play the major role in magnetic energy dissipation and thus constitute an important factor in determining the global temperature of the solar wind.

\section{Summary and conclusions}\label{sec:Summary}
 In this work, we have analyzed magnetic field and particle data from the first six encounters of the PSP mission. Our goal was to study the statistics of intermittency and further elucidate the nature of turbulent dissipation in the neighborhood of the solar wind sources. As a first step, an effort was made to understand the nature of the mechanism that is responsible for the generation of intermittency and coherent structures in the solar wind. We have shown that coherent structures, corresponding to $PVI \geq 1$, constitute only $\approx 19 \%$ of the dataset. As a follow-up to the study of  \cite{Chhiber_2020}, we studied the waiting time distributions by applying thresholds on the PVI time-series. We have confirmed that intermittent magnetic field structures are not evenly distributed in the solar wind but rather tend to strongly cluster, forming regions characterized by a high magnetic field variability followed by intervals for which the magnetic field is relatively smooth. This observation is also reinforced by the power-law nature of the waiting time distributions at low waiting times, indicating the presence of an intracluster population. The power law is followed by an exponential at longer waiting times, suggesting a second intercluster population of coherent structures in the solar wind's magnetic field. Additionally, the power-law scaling of the WT distributions is indicative of clusters that do not have a typical size or distance, except for the limiting size given by the exponential cutoff.
  We moved on to investigate the correlation of identified coherent structures to the proton temperature as a proxy to intermittent dissipation in the solar wind. In Sec. \ref{subsec:heating}, we have provided strong evidence that supports the theory of plasma heating by coherent magnetic structures generated by the turbulent cascade. More specifically, in agreement with previous studies \citep{2012PhRvL.108d5001S, 2020ApJS..246...46Q, 2021ApJ...921...65Y}, the present analysis suggests that the strongest discontinuities, $PVI \geq 6$ in the magnetic field, are linked to extreme dissipation events. In the past, both theoretical (e.g., \cite{Servidio11}) and observational studies (e.g., \citep{2021ApJ...908..237H}), have demonstrated that such events are most likely to be associated with magnetic reconnection sites.  A visual inspection of tens of high PVI valued events recovered during $E_{1}-E_{6}$ supports this theory, as in most cases, such events are related to reversals of the magnetic field and zero-crossings of at least one of the magnetic field components, as well as significant energy dissipation and proton heating. We have demonstrated in our analysis that such high PVI value events can considerably alter the local plasma dynamics by increasing $T_{p}$ in a very short amount of time. However, as indicated by Fig. \ref{fig:heating5}, and,  Fig. \ref{fig:heating6}, due to the relatively small fraction of these structures in our dataset, their contribution in energy dissipation is practically insignificant. Consequently, it is rather the accumulative effect of many small-scale coherent structures of PVI index in the $1 \leq PVI \leq 6$ range that determines $T_{p}$ on a macroscopic scale. This means  $T_{p}$ is strongly related to the number density of coherent structures in our dataset, with high-density intervals being associated with higher $T_{p}$. In accordance with our study, \cite{2021ApJ...908..237H}, have recently analyzed measurements from the Magnetospheric Multiscale Spacecraft (MMS) mission to show that despite having a significant role in local dissipation, magnetic reconnection events are not the major contributor to the magnetosheath's internal energy. This study provides an additional confirmation that the number density of intense coherent structures such as reconnection sites and current sheets is an important parameter that should be taken into account in numerical investigations that study the heating of astrophysical and space plasmas. \par
  Finally, it would be important to compare our results to those of previous studies investigating intermittent heating at different heliospheric distances. This comparison can allow us to infer the relative importance of intermittent heating on different phases of the development of the turbulent cascade. Such an example is Fig. 3 of \cite{PhysRevLett.108.261102}, who studied the effects of intermittent dissipation on $T_{p}$ at 1AU. The caveat here is, that \cite{PhysRevLett.108.261102} have considered values of $PVI \geq 2.4$ to interpret $T_{p}$ as function of $f_{PVI \geq 2.4}$. However, even though not clearly outlined in \cite{PhysRevLett.108.261102}, events of $1 \leq PVI \leq 2.4$ seem to also contribute to the heating of the solar wind protons (e.g., see  Fig. 4).  Nevertheless out of this comparison, it is clear that the effects of intermittent heating on $T_{p}$ close to the solar wind sources are not as pronounced as in the outer parts of the heliosphere. A possible explanation for the decreased efficiency of intermittent heating in the young solar wind is that PSP mostly samples intervals for which the angle between the background magnetic field and the solar wind flow $\Theta_{VB}$ is parallel.  
  
In the past, intermittency in the inertial range of MHD turbulence has been shown to be highly anisotropic. For parallel intervals, the statistical signature of the magnetic field fluctuations is that of a non-Gaussian globally scale-invariant process, in contrast to multi-exponent statistics observed when the local magnetic field is perpendicular to the flow direction  \citep{2008PhRvL.101q5005H, 2014AGUFMSH52A..03O}. This result can be interpreted as a decreased level of intermittency for parallel intervals and consequently a decreased efficiency of the intermittent heating mechanism. We will soon revisit the topic by studying the evolution of intermittent heating as a function of helispheric distance, as well as $\Theta_{VB}$ (Sioulas et al. in preparation).

As noted in Sec. \ref{sec:intro}, the advantage of the PVI method to capture all kinds of discontinuities can also be its biggest drawback. Namely, two equally valued PVI events may correspond to different types of magnetic field discontinuities, that have contrasting effects on the proton temperature. For example, different types of coherent structures may instigate heating mechanisms that preferentially operate close but not inside magnetic field discontinuities. This could strongly affect the results of this study, as heating may not coincide in time with strong discontinuities in the magnetic field. Thus to further categorize such discontinuities in the near sun solar wind environment, several additional magnetic field and plasma parameters should be taken into account (e.g.,  \cite{article}). We hope that future studies will revisit the topic by performing a more thorough analysis, considering such parameters.

\vspace*{0.1cm}

\begin{acknowledgements}

We would like to thank the anonymous referee for their insightful suggestions and comments, which helped improve and clarify this manuscript. \par
This research was funded in part by the FIELDS experiment
on the Parker Solar Probe spacecraft, designed and developed under NASA contract
NNN06AA01C; the NASA Parker Solar Probe Observatory Scientist
grant NNX15AF34G and the  HERMES DRIVE NASA Science Center grant No. 80NSSC20K0604. 

\end{acknowledgements}

\bibliography{sample631}{}
\bibliographystyle{aasjournal}



\end{document}